\documentclass[10pt, twocolumn]{article}
\usepackage[dvips]{graphicx}
\usepackage{color}
\usepackage{epsfig}
\usepackage{cite}
\usepackage{amsmath,latexsym,mathrsfs}
\def\BA{\begin{eqnarray}} \def\BE{\begin{equation}}
\def\EA{\end{eqnarray}} \def\EE{\end{equation}} 
 
\def\gtsim{\lower-0.45ex\hbox{$>$}\kern-0.77em\lower0.55ex\hbox{$\sim$}}
\def\ltsim{\lower-0.45ex\hbox{$<$}\kern-0.77em\lower0.55ex\hbox{$\sim$}}
\newcommand{\ir}{\text{\tiny IR}}

\newcommand{\qcd}{\text{\tiny QCD}}
\newcommand{\ym}{\text{\tiny YM}}

\begin{document}
\title{Equivalence of the AdS-Metric and the QCD Running Coupling} 
\author{H.~J.~Pirner$^{ab}$\thanks{pir@tphys.uni-heidelberg.de}\ ,\
B. Galow$^{ab}$\\ {}\\ ${}^a$Institut f\"ur Theoretische Physik
der Universit\"at Heidelberg, Germany\\ ${}^b$Max-Planck-Institut
f\"ur Kernphysik Heidelberg, Germany} \maketitle
\begin{abstract}
\noindent 
We use the functional form of the QCD running coupling to modify the conformal metric in
AdS/CFT mapping the fifth-dimensional z-coordinate to the energy 
scale in the four-dimensional QCD. The resulting type-0 string theory in
five dimensions is solved with the Nambu-Goto action giving good agreement
with the 
Coulombic and confinement  $Q \bar Q$ potential. 

%\vspace{1pc}
\end{abstract}

%\begin{document}

\section{Introduction}

The AdS/CFT conjecture relates type IIB superstring theory in the
AdS$_5\times$S$^5$
background with four-dimensional super Yang Mills
theory. Supersymmetric QCD is scale invariant with a vanishing
$\beta$-function. In contrast, QCD has 
no supersymmetry and a non-vanishing $\beta$-function with a well
defined running coupling. This defines in our opinion the first task of
how to modify the background of supergravity on AdS$_5\times$S$^5$, in order to
obtain a more QCD-like theory: We have to break conformal invariance
and disregard supersymmetry. One possible coordinate system of AdS$_5$ is
given in the near horizon limit by
\BE
ds_{\text{\tiny{radial}}}^2=\frac{L^2}{r^2}dr^2+\frac{r^2}{L^2}(-dt^2+d\vec{x}
^2).
\label{horizonmetr}
\EE

In the radial coordinates the boundary of the space is at $r\to\infty$.
We are going to use another coordinate system,
the Poincar\'e
coordinates, related with the radial coordinates by the transformation
$z=L^2/r$. These coordinates are also called conformal coordinates, because
one can directly read off the scale invariance of the metric in this coordinate
patch

\BE
ds^2=G_{\mu\nu} dX^\mu dX^\nu=\frac{L^2}{z^2}(-dt^2+d\vec{x}^2+dz^2).
\label{Adsmetr}
\EE

$L$ is the radius of AdS$_5$. The boundary of the AdS space is at $z=0$. A
simple ansatz of breaking conformal invariance is to multiply the
metric in eq. (\ref{Adsmetr}) by the so-called warping function. One
can show \cite{Randall:1999ee} that global Poincar\'e invariance demands that
the warping function has to depend only on the $z$-coordinate. Thus the new
metric is of the form

\BE
ds_{\qcd}^2=h(z)\cdot ds^2=h(z) \frac{L^2}{z^2}(-dt^2+d\vec{x}^2+dz^2),
\label{QCD_Metr}
\EE

where the subscript $QCD$ symbolizes that this is not anymore a metric for the
AdS$_5$ space but is a first attempt to obtain a QCD-like theory from this
modified AdS$_5$ space.\\
How should one choose $h(z)$? QCD is a renormalizable
quantum field theory. Hence, the UV divergences can be
absorbed in a renormalized 
coupling $g$.  It is common to use the strong coupling constant 
$\alpha_s=g^2/4\pi$. This coupling is given to the lowest order by the
formula

\BE
\alpha_s(p)=\frac{1}{4\pi\beta_0\log(p^2/\Lambda_{\qcd}^2)}=\frac{4\pi}{
(11-\frac{2}{3}n_f)\log(p^2/\Lambda_{\qcd}^2)}.
\label{running_coupling}
\EE

Here $p$ is the scale, which can be chosen to coincide with
the  transferred
momentum \cite{Yndurain:1999ui}. $\Lambda_{\qcd}$ is called the QCD
scale parameter,
which is to be
determined by experiments. Finally
$\beta_0=\frac{1}{(4\pi)^2}(11-\frac{2}{3}n_f)$ is the absolute 
value of the first coefficient of the $\beta$-function, which is
subtraction-scheme-independent. One can see that, for
$p\to\infty$, the coupling $\alpha_s$ vanishes, and the theory becomes scale
invariant.\\

It can be shown \cite{Aharony:1999ti} that the radial coordinate
$r$ and thus also the coordinate $z$
corresponds to the energy scale  $p\propto 1/z$ of the boundary field theory.
Hence, we have for small values of $z$ the UV region of the boundary field
theory, and for large values of $z$ we are in the IR region of the boundary
field theory.
Therefore, the bulk space contains all possible energy scales of the boundary
field theory \cite{Kiritsis:2007zz}. For QCD correlation functions, it is
natural to have
a scale-invariant theory in the limit $z\to 0$. In that limit, we should have
$h(z)\to 1$.
Given the fact that the warp factor and the dilaton (which determines the
running coupling) are not independent and coupled via the 5D Einstein
equations, it seems reasonable to try an ansatz for the warping function that
equates it with $\alpha_s$ given in (\ref{running_coupling}). We will see in the
rest of the paper
that such an ansatz leads to very good agreement with the Cornell potential.

\BE
h(z)=\frac{c_2}{\log\left [\frac{1}{z^2+l_s^2}\frac{1}{\Lambda^2}\right ]}.
\label{hz_general}
\EE

The introduction of the parameter $l_s$
guarantees the conformal limit
$h(z)\to \text{finite}$ at $z\to 0$. The
requirement $h(z)\to 1$
for $z\to 0$ fixes 
$c_2=\log\left (\frac{1}{l_s^2\Lambda^2}\right )$.
We  assume that
$\Lambda$ is related to the AdS$_5$-radius as
$\Lambda=1/L$ and define the dimensionless parameter $\epsilon$ as the ratio

\BE
\epsilon\equiv\frac{l_s^2}{L^2}= l_s^2 \Lambda^2.
\label{small}
\EE 

Then we obtain
\BE
h(z)=\frac{\log\left (\frac{1}{\epsilon} \right )}{\log\left
[\frac{1}{(\Lambda z)^2+\epsilon}\right ]},
\label{hz_log}
\EE

with the IR singulartiy at

\BE
z_\ir=\sqrt{\frac{1-\epsilon}{\Lambda^2}}.
\EE
According to the equivalence of $1/z$ to energy resolution in
four dimensions
the scaling factor $h(z)$ indirectly encodes the
running behavior of the strong coupling $\alpha_s$.
\\
In the infrared we have broken the conformal invariance 
by cutting off the AdS$_5$ space at some finite value of $z=z_{\ir}$.
In the language of AdS/QCD
this set-up is very similar to a hard-wall model
\cite{Polchinski:2001tt}. 
As we will demonstrate the modified coupling naturally incorporates
confinement at the infrared energy scale $\Lambda\approx 1/z_\ir$. 
In the ultraviolet QCD becomes a scale invariant theory at
sufficiently high energies because of asymptotic freedom, which shows
up in the metric preserving the conformal form, i.e. $\mathop
{\lim }\limits_{z \to 0}h(z)=1$.\\

%%%%%%%%%%%%%%%%%%%%%%%%%%%%%%%%%%%%%%%%%%%%%%%%%%%%%%%%%%%%%%%%%%%%%%%%%%%%%

\section{Heavy quark potential from AdS/QCD}

\begin{figure}[h!]
\begin{center}
\begin{picture}(0,0)%
\includegraphics{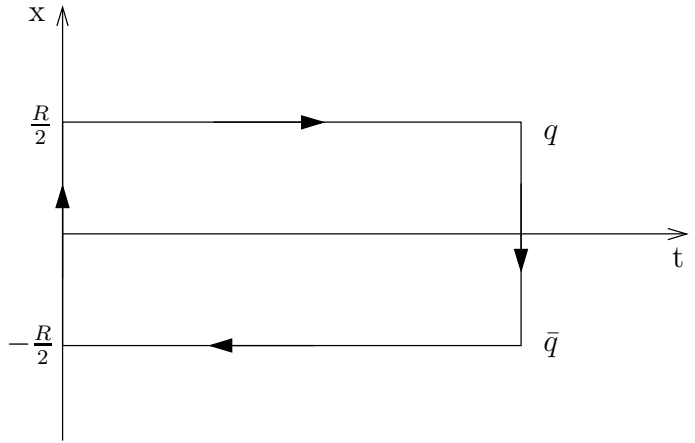}%
\end{picture}%
\setlength{\unitlength}{4144sp}%
\begingroup\makeatletter\ifx\SetFigFont\undefined%
\gdef\SetFigFont#1#2#3#4#5{%
  \reset@font\fontsize{#1}{#2pt}%
  \fontfamily{#3}\fontseries{#4}\fontshape{#5}%
  \selectfont}%
\fi\endgroup%
\begin{picture}(3993,2634)(5131,62)
\put(8242,1881){\makebox(0,0)[lb]{\smash{{\SetFigFont{12}{14.4}{\familydefault}{
\mddefault}{\updefault}{\color[rgb]{0,0,0}$q$}%
}}}}
\put(8242,609){\makebox(0,0)[lb]{\smash{{\SetFigFont{12}{14.4}{\familydefault}{
\mddefault}{\updefault}{\color[rgb]{0,0,0}$\bar q$}%
}}}}
\put(5030,600){\makebox(0,0)[lb]{\smash{{\SetFigFont{12}{14.4}{\familydefault}{
\mddefault}{\updefault}{\color[rgb]{0,0,0}$-\frac{R}{2}$}%
}}}}
\put(9012,1111){\makebox(0,0)[lb]{\smash{{\SetFigFont{12}{14.4}{\familydefault}{
\mddefault}{\updefault}{\color[rgb]{0,0,0}t}%
}}}}
\put(5164,1914){\makebox(0,0)[lb]{\smash{{\SetFigFont{12}{14.4}{\familydefault}{
\mddefault}{\updefault}{\color[rgb]{0,0,0}$\frac{R}{2}$}%
}}}}
\put(5164,2584){\makebox(0,0)[lb]{\smash{{\SetFigFont{12}{14.4}{\familydefault}{
\mddefault}{\updefault}{\color[rgb]{0,0,0}x}%
}}}}
\end{picture}%
\end{center}
\caption{Rectangular Wilson loop contour put on the 4-dimensional boundary of
the modified AdS$_5$ space.}
\label{Wilson_loop_Contour}
\end{figure}

In QCD one can include quarks as infinitely heavy external
probes.
To determine the interaction potential $V_{Q\bar Q}(R)$ between the quark and
the antiquark we use 
the Wilson loop. The Wilson loop describes the creation of a $Q\bar Q$-pair at
some time t$_1$, interaction of the created quark and the antiquark with 
themselves and the vacuum during a period of time $T$, and the
annihilation of the pair at time t$_2$. 
The Wilson loop, c.f. Fig. \ref{Wilson_loop_Contour}

is defined as

\BE
W[C]=\frac{1}{N} Tr P \exp[i \oint_{C} A_\mu dx^\mu].
\label{Wilson_loop_formula}
\EE

The $1/N$-factor is introduced for convenience because there are $N$ terms in
the trace over 
the unit matrix in the fundamental representation of an $SU(N)$ gauge theory,
and the $P$ stands for
path ordering of the exponential:
For $T\to\infty$,  the VEV of the Wilson loop
behaves as $\langle W(C)\rangle\propto
e^{-T V_{Q\bar Q}}$.\\

According to the
holographic dictionary \cite{Witten:1998qj,Gubser:1998bc}, the expectation value
of the Wilson
loop in four dimensions should be equal to the string partition function on the
modified AdS$_5$ space, with the string worldsheet ending
on the contour C at the boundary of AdS$_5$

\BE
\langle W^{\text{4d}}[C]\rangle=Z_{\text{string}}^{\text{5d}}[C]\approx
e^{-S_{NG}[C]}.
\EE

The second relation is obtained by the saddle-point approximation, in which the
partition function is just given by the classical action
\cite{Maldacena:1998im}. Hence, we have to consider
the classical string worldsheet action $S_{NG}$. As in the original
hadronic string theory the Nambu-Goto action will play a major role to
model the gluonic degrees of freedom. However, the string-gauge
theory has to be extended to gravity if one looks for a
consistent explanation of the metric as a solution of the Einstein
equations with a dilaton. This work will be published separately. 
Note the string worldsheet is embedded into the
five-dimensional bulk space. The worldsheet is stretching from the boundary
of AdS$_5$ at infinity down to a given point resulting in an infinite worldsheet
area and thus $\langle W[C]\rangle=0$. Since our worldsheet is swept out by an
infinitely
heavy string, the mass of the string times the length
of the loop $C$ should be subtracted
from $S_{NG}$ \cite{Maldacena:1998im,Kiritsis:2007zz}. The
resulting difference is finite. This is incorporated in the 
later performed UV renormalization of the Nambu-Goto action.\\

To calculate the $Q\bar Q$ potential we use the AdS/QCD background
Euclidean metric

\BE
ds_{Eucl}^2=G_{\mu\nu} dX^\mu dX^\nu=\frac{h(z) L^2}{z^2}(dt^2+d\vec{x}^2+dz^2).
\label{eucl_Adsmetr}
\EE

The Nambu-Goto action $S_{NG}$ is given by

\BE
S_{NG}=\frac{1}{2\pi l_s^2}\int d^2\xi\sqrt{\det h_{ab}},
\EE

where $l_s$ is the string
length and 
$h_{ab}$ is the induced worldsheet metric:
The indices ${a,b}$ are reserved to the $\xi_1,\xi_2$-coordinates on
the worldsheet, the greek indices ${\mu,\nu}$ to the coordinates of
the embedding five-dimensional space

\BE
h_{ab}=G_{\mu\nu}\frac{\partial X^\mu}{\partial\xi^a}
\frac{\partial X^\nu}{\partial\xi^b}.
\EE

In the static gauge, the worldsheet coordinates can be chosen as $\xi^1=t$ and $\xi^2=x$.
In such a static configuration $z=z(x)$ is the only
$x$-dependent function. The Wilson-loop contour $C$ is located at the boundary
of the AdS space, i.e. at $z\to 0$. The set-up is presented in Fig.
\ref{Wilson_loop_Contour}.

The induced worldsheet metric obtained from eq. (\ref{eucl_Adsmetr})

\BE
h_{ab}=\frac{L^2 h(z)}{z^2}\left( \begin{array}{cc} 1 & 0 \\ 0 &
1+(\frac{\partial z}{\partial x})^2 \end{array} \right)
\label{habGeneral}
\EE

has to be put into the Nambu-Goto action together with the dimensionless
parameter $\epsilon=\frac{l_s^2}{L^2}$ to get 

\BE
S_{NG}=\frac{T}{2 \pi \epsilon} \int dx \frac{h(z)}{z^2}\sqrt{1+(z')^2},
\label{sCoulomb}
\EE

where $z'=\frac{dz}{dx}$ and T comes from the integral over time.
Now we can identify

\BE
\mathscr{L} (z,z')=\frac{h(z)}{z^2}\sqrt{1+(z')^2}
\label{LagrangianCoul}
\EE

with an effective Lagrangian, and the problem reduces to a simple problem of
classical mechanics with the
Hamiltonian

\BE
\mathscr{H}=p(z)\cdot z'-\mathscr{L},
\EE

where $p(z)=\frac{\partial\mathscr{L}}{\partial z'}$ is the conjugate momentum.
One obtains

\BE
\mathscr{H}=\frac{-h(z)}{z^2\sqrt{1+(z')^2}}.
\EE

Energy conservation allows one to set $\mathscr{H}=-1/c^2$, where $c$ is a
constant

\BE
\frac{h(z)}{z^2\sqrt{1+(z')^2}}=\frac{1}{c^2}.
\label{hcond}
\EE

We express this integration constant $c$ via the
maximal value of $z$, which we denote as $z_0$.
Equation (\ref{hcond}) yields at $x=0$:

\BE
\frac{h(z_0)}{z_0^2}=\frac{1}{c^2}.
\label{maxcond}
\EE

We can rewrite eq. (\ref{hcond}) as:

\BE
z'=\sqrt{\left (\frac{h(z) c^2}{z^2}\right )^2-1}.
\EE
Using the condition eq. (\ref{maxcond}) for the maximum and rescaling
$z=\nu z_0$, we obtain the inter-quark distance $R$ as a function of $z_0$:

\BE
R(z_0)=2 z_0 \int_0^1 d\nu \nu^2 \frac{h(z_0)}{h(\nu z_0)}\frac{1}{\sqrt{1-\nu^4
\left (\frac{h(z_0)}{h(\nu z_0)}\right )^2}}.
\label{distance}
\EE

By similar transformations we can write the energy, which we get from the
Nambu-Goto string action, as a function of $z_0$

\BE
V_{Q\bar Q}(z_0)=\frac{1}{\pi\epsilon}\frac{1}{z_0}\int_0^1 d\nu \frac{h(\nu
z_0)}{\nu^2}\frac{1}{\sqrt{1-\nu^4\left ( \frac{h(z_0)}{h(\nu z_0)}\right )^2}}.
\label{energy}
\EE

The regularization of the potential is related to the subtraction of the masses 
of infinitely heavy
quarks as discussed before.
We subtract the singular part $\propto \frac{1}{\nu^2}$ from the integrand and add its
primitive at the upper limit which results in

\BE
\begin{split}
V_{Q\bar
Q}^{ren.}(z_0)=-\frac{1}{\pi\epsilon}\frac{1}{z_0}
+\frac{1}{\pi\epsilon}\frac{1}{z_0}\int_0^1 d\nu
\left (\frac{h(\nu
z_0)}{\nu^2}\vphantom{\frac{1}{\sqrt{1-\nu^4 \left( \frac{h(z_0)}{h(\nu z_0)}
\right)^2}}-\frac{1}{\nu^2}}\right.\\ \left.
\frac{1}{\sqrt{1-\nu^4\left ( \frac{h(z_0)}{h(\nu z_0)}\right
)^2}}-\frac{1}{\nu^2}\right ).
\end{split}
\EE

\begin{figure}[h!]

\begin{picture}(3,3)
\put(5,-8){$R(z_0)[GeV^{-1}]$}
\put(167,-135){$z_0[GeV^{-1}]$}
\end{picture}

\begin{center}
\includegraphics[width=0.45\textwidth]{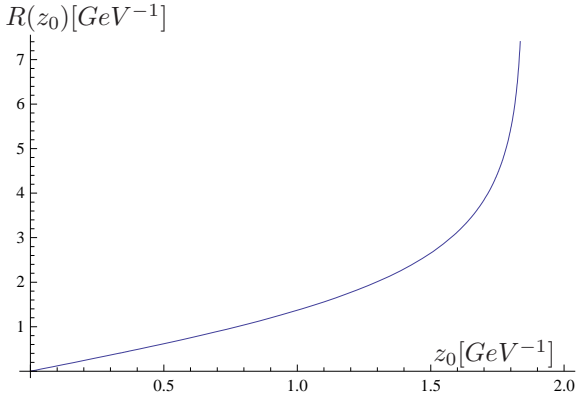}
\end{center}
\caption{A plot showing $R(z_0)$ for $\epsilon=0.48$ and
$\Lambda=0.264\,\mbox{GeV}$.}
\label{Rofz0}

\end{figure}

\begin{figure}[h!]

\begin{picture}(3,3)
\put(5,-4){$V_{Q\bar Q}(z_0)[GeV]$}
\put(170,-64){$z_0[GeV^{-1}]$}
\end{picture}
\begin{center}
\includegraphics[width=0.45\textwidth]{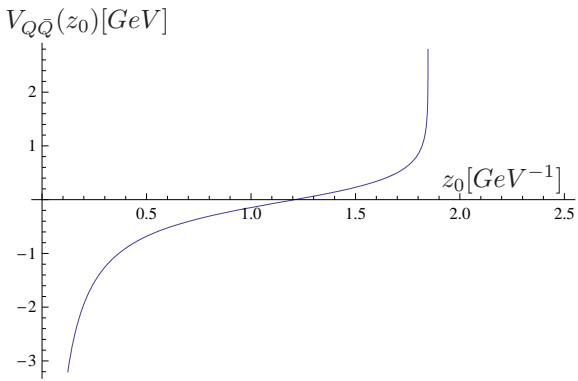}
\end{center}
\caption{A plot showing $V_{Q\bar Q}(z_0)$ for $\epsilon=0.48$ and
$\Lambda=0.264\,\mbox{GeV}$.}
\label{Vofz0}

\end{figure}

We continue evaluating eqs. (\ref{distance}) and (\ref{energy}) in
terms of the parameter $z_0$. In order to get a first impression how both
integrals depend on $z_0$, we plot them in Figs. \ref{Rofz0} and 
\ref{Vofz0}.

The phenomenological Cornell potential of the form $V_{Q\bar
Q}=-\frac{a}{R}+\sigma R$ determines the parameters
in the underlying metric. 
We fix the dimensionless parameter $\epsilon$ to the parameter $a$ in the
Coulombic part of the potential and the 
parameter $\Lambda$ to the
string tension $\sigma$ in the long distance $Q \bar Q$-interaction. 
It is quite natural to have two parameters in the metric to determine
two parameters in the potential. The agreement with the
phenomenological potential can be improved \cite{Andreev:2006ct}. Indeed the
found $\Lambda$ will be similar to 
$\Lambda_{\overline{\text{\tiny{MS}}}}$ in QCD.
The QCD string has been the origin of hadronic string theory which has
been supported by lattice simulations of QCD where one identifies the stretched
tube of the color
electric flux with a string.
We can see from Fig. \ref{Rofz0} that the $Q\bar Q$ distance $R$ depends
linearly on $z_0$ for small $z_0$. Looking at the definition of $h(z)$ in eq.
(\ref{hz_log}) we realize that, for $z_0\approx 0$, the $\nu$-dependence of
$h(z)$ is suppressed. Hence, we only make a negligible error when performing
Taylor expansion of the integrand of $R(z_0)$ in eqs. (\ref{distance}) and
(\ref{energy}) at $z_0=0$ up to the first order and then integrating over $\nu$
in order to obtain the
behavior of the potential $V_{Q\bar Q}$ at small $Q\bar Q$-separations $R$. For
eq. (\ref{distance}) this yields

\BE
R(z_0)=2\sqrt{\pi}\frac{\Gamma(3/4)}{\Gamma(1/4)}z_0+\mathcal{O}(z_0^2),
\label{Rofsmallz0}
\EE

which is exactly the result we would have obtained in the
conformal case with $h(z)=1$.

We expand the integrand of eq. (\ref{energy}) at $z_0=0$ to the order
$\mathcal{O}(z_0^2)$, integrate over $\nu$ and then insert $z_0(R)$ 
from eq. (\ref{Rofsmallz0}), and finally obtain

\BE
V_{Q\bar Q}(R)=-2\left(\frac{\Gamma(3/4)}{ 
\Gamma(1/4)}\right)^2\frac{1}{\epsilon R}+\sigma R,
\label{small_Potential}
\EE
with
\BE
\begin{split}
\sigma
=\left(-\frac{1}{4\pi} + \frac{27}{256\pi
}\frac{ \Gamma(1/4)^2}{\Gamma(7/4)^2}\right)\frac { \Lambda^2 } {
\epsilon^2\log ( 1/\epsilon) }\\\approx0.443\frac { \Lambda^2 } {
\epsilon^2\log ( 1/\epsilon) }.
\end{split}
\label{small_tension}
\EE

We can see from the $1/R$-term in eq. (\ref{small_Potential}) that it is
exactly the same as in the case of the conformal metric 
\cite{Maldacena:1998im}. In order to adjust the above
potential to the value of the Coulombic
part of the $Q\bar Q$ interaction  given in
\cite{Eichten:1978tg}, namely $V_{Q \bar
Q}=-\frac{a}{R}+\sigma R$ with $a=0.48$ and $+\sigma=0.183\,\text{GeV}^2$ we
have to choose 

\BA
\epsilon&=& 0.48 \\
\Lambda&=&264\,\text{MeV}.
\EA

\begin{figure}[t!]

\begin{picture}(3,3)
\put(4,-5){$V_{Q\bar Q}^{ren.}(R)[GeV]$}
\put(169,-66){$R[GeV^{-1}]$}
\end{picture}

\begin{center}

\includegraphics[width=0.45\textwidth]{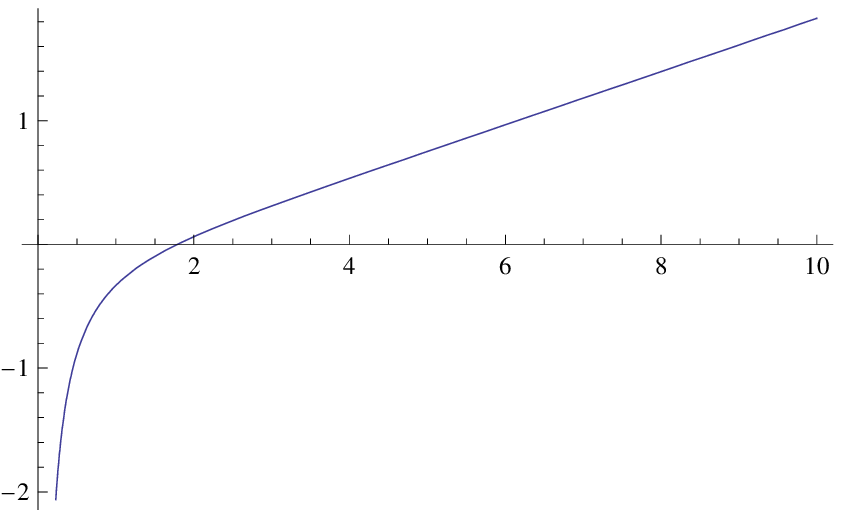}
\end{center}
\caption{Numerically calculated heavy-quark potential for the modified metric,
eq. (\ref{eucl_Adsmetr}), using $h(z)$ from eq. (\ref{hz_log}).}
\label{CornellPotential}
\end{figure}

This result looks rather reasonable. For example, the value of the scale
parameter in four-flavor QCD\footnote{We do not want to compare exactly to four-flavor
QCD, but want to show that $\Lambda$ has the correct magnitude.} is
$\Lambda_{\qcd}^{n_f=4}=274\pm30\,\text{MeV}$
\cite{Yndurain:1999ui}. Having fixed the two parameters we can now numerically
evaluate the heavy quark potential to test the form on all length scales. 
Fitting the numerical potential  plotted  in the interval
$R\in[0.1\,\text{GeV}^{-1},9.6\,\text{GeV}^{-1}]$  in Fig.
\ref{CornellPotential} to a
Cornell-like potential $V_{Cornell}(R)=-\frac{a}{R}+\sigma R$ yields
$a=0.47$, $\sigma=0.181\,\text{GeV}^2$. If one takes
the dependence of these parameters on the fit intervall into account
the numerically determined values coincide with the analytical ones. \\

For further applications it is important to note 
that the validity of a gravity dual to the string description 
is $\frac{L^4}{l_s^4}\gg1$. We
obtain $\frac{L^4}{l_s^4}\approx 4.3$. This choice is imposed by the
$Q\bar Q$ potential. So it may be necessary to include higher
correction to the simple form of gravity with an 
AdS-negative cosmological constant. One can try to
include these corrections via a modified dilaton potential in the corresponding
five-dimensional gravity theory. Due to the form of the warp factor, the
resulting dilaton dynamics may only reproduce the
$\beta$-function of QCD approximately. \cite{newpaper}

In perturbative QCD one knows that for $R\to 0$ the potential $V_{Q\bar
Q}=-\frac{N_c^2-1}{2N_c}\alpha_s\frac{1}{R}$ and hence scales with $g_{\ym}^2
N_c$. It is possible to show that
$\frac{L^4}{l_s^4}=g_{\ym}^2 N_c$ \cite{Kiritsis:2007zz}. The potential
obtained, eq.
(\ref{small_Potential}), is
only proportional to
$\frac{L^2}{l_s^2}$ and hence scales with $g_{\ym} \sqrt{N_c}$. An explanation
of this discrepancy may be that the limit $l_s\to0$ 
corresponds to a Yang-Mills theory with a strong coupling $g_{\ym}^2 N_c$. For
realistic AdS/QCD this argument has to be studied in more detail.
An interesting result of our calculation is the
proportionality between 
$\sigma$ and $\Lambda$ given by eq. (\ref{small_tension}), once
$\epsilon$ is fixed.\\

From the taylor expansion in eq. (\ref{sing_exp}) one can see that there exists
a complex singularity at $z_0=z_{\gamma}$.
Let us determine  $z_{\gamma}$. From eq. (\ref{distance}), one can see
that the dominant contribution to the integral arises at $\nu=1$. A Taylor
expansion of the integrand at $\nu=1$ followed by the integration yields
up to $\mathcal{O}(1-\nu)$ the following expression

\BE
R(z_0)=\frac{2 i z_0}{\sqrt{-1+\frac{z_0^2
\Lambda ^2}{\left(\epsilon+z_0^2 \Lambda
^2\right) \text{Log}\left[\frac{1}{\epsilon+z_0^2 \Lambda ^2}\right]}}},
\label{sing_exp}
\EE 
 
which is only real for

\BE
\frac{z_0^2
\Lambda ^2}{\left(\epsilon+z_0^2 \Lambda
^2\right) \text{Log}\left[\frac{1}{\epsilon+z_0^2
\Lambda^2}\right]}<1.
\EE

This inequality can be solved in terms of the ProductLog
function\footnote{$\text{ProductLog(f)}$ gives the principal solution for $w$
in $f=w e^{w}$.}, and one obtains for the above determined parameters

\BA
z_0&<&z_{\gamma}\\
z_{\gamma}&=&\sqrt{\frac{\epsilon}{\Lambda^2}\left
(\frac{1}{\text{ProductLog}(\epsilon
e )}-1\right )}\nonumber\\ &\approx& 1.85\,\text{GeV}^{-1},
\EA

where the $e \approx 2.71828$ in the denominator is the base of the natural
logarithm. A similar analysis of
the integral in eq. (\ref{energy}) yields the same complex singularity at
$z_0=z_\gamma$. 

This
singularity defines a 
horizon\footnote{This horizon should not be confused with
the IR singularity of the modified metric at
$z_\ir\approx2.73\,\text{GeV}^{-1}$.} in contrast
to the purely conformal AdS background, where there
is
no upper bound on the parameter $z_0$. For larger $R$-values, one obtains
a larger $z_0$. In case of the modified metric, the $z_0$ of the
worldsheet is limited by the
horizon. For other confining backgrounds 
based on the running coupling we refer to
the studies of Kiritsis et al. in \cite{Gursoy:2007cb, Gursoy:2007er}. In these
papers,
the authors analyze various confining backgrounds by studying the long-range
part of
the $Q\bar Q$-potential, given in the form derived in \cite{Kinar:1998vq}. It
should be mentioned that our background given by $h(z)$ of eq. (\ref{hz_log})
satisfies the criterium for a confining background by eq. (3.12) of Ref.
\cite{Gursoy:2007er}. We also refer to \cite{White:2007tu}, where the
Cornell potential is derived in various backgrounds.\\ 

\begin{figure}[t!]

\begin{picture}(3,3)
\put(-25,-69){$z[GeV^{-1}]$}
\put(120,-147){$x[GeV^{-1}]$}
\end{picture}
\begin{center}

\includegraphics[width=0.45\textwidth]{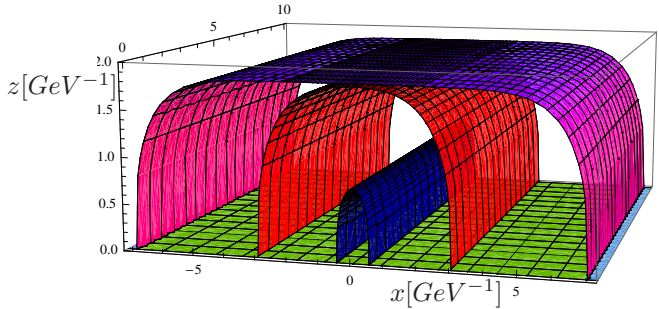}
\end{center}
\caption{Nambu-Goto worldsheet for the non-conformal metric of eq.
(\ref{eucl_Adsmetr}) at the values of
the $Q\bar Q$-separation $R=1\,\text{GeV}^{-1}$ (blue), $R=6\,\text{GeV}^{-1}$
(red), and $R=14\,\text{GeV}^{-1}$ (magenta) plotted over the Wilson loop
area (green) in the $(x,t)$ plane.}
\label{BulkMotionNonConformal3d}
\end{figure}

Finally we show the minimal worldsheets for
the modified metric given by eq. (\ref{eucl_Adsmetr})
in Fig. \ref{BulkMotionNonConformal3d}.\\

One can see that the worldsheets are
flattening with the increase of the $Q\bar Q$-separation $R$.
There exists one crucial difference compared to the
conformal case,
namely that we have an upper bound $z_{\gamma}\approx1.85\,\text{GeV}^{-1}$,
dictated by the complex singularity
discussed above. An upper bound on $z_0$ implies an upper bound for $R$. For
the parameter values $\epsilon=0.48$ and $\Lambda=0.264\,\text{GeV}$, we obtain
for the upper bound $R\approx 2.8\,\text{fm}$. We can see from Fig.
\ref{BulkMotionNonConformal3d} that the
worldsheet already reaches its maximal $z$-value for $x$ away from zero, and the surface
becomes completely flat, which makes a confining potential. One can
interpret this effect as touching the IR 
horizon. Dubovsky and Rubakov \cite{Dubovsky:2001pe} obtain a quite similar
behavior, while studying two charges in front of a brane. There, the electric
flux of the charges drops down on the brane, while having still confinement.
We do not have string-breaking effects, despite
having a maximal $Q\bar Q$-separation $R\approx 14.1\,\text{GeV}^{-1}$. This is
due to the fact that we have no dynamical quarks.

%%%%%%%%%%%%%%%%%%%%%%%%%%%%%%%%%%%%%%%%%%%%%%%%%%%%%%%%%%%%%%%%%%%%%%%%%%%%%

\section{Conclusion}

We have chosen the functional form of the warping factor of the AdS-metric eq.
(\ref{hz_log}) such that it coincides with the functional form of the QCD
running coupling eq.
(\ref{running_coupling}). Using the holographic dictionary
\cite{Witten:1998qj,Gubser:1998bc,Maldacena:1998im} we then extract
the $Q\bar Q$-potential. Only one new parameter $\epsilon=0.48$ has to be
fitted  
to reproduce the short and long range 
Cornell potential  \cite{Eichten:1978tg}. The other parameter
$\Lambda=264\,\text{MeV}$ in the metric
is close to $\Lambda_{\qcd}^{n_f=4}$. The phenomenology of equivalence
proves to be
successful and simple. Comparing with one of the previous calculations 
 \cite{Andreev:2006ct} of the  $Q\bar Q$-potential one sees in Tab.
\ref{AndreComp}
that our work can reproduce both the strength of the Coulomb interaction $a$ and the string
tension $\sigma$ at the same time. The parameter $L^4/l_s^4\gg1 $ indicates that
the string
theory has a meaningful
gravity approximation which we will present in a separate paper \cite{newpaper}.

\begin{center}

\begin{table}[t!]

\renewcommand{\arraystretch}{1.5}

\begin{center}
\begin{tabular}{|c|c|c|c|}

\hline
 & \cite{Andreev:2006ct} & our work & Cornell
\cite{Eichten:1978tg} \\
\hline
$L^4/l_s^4$ & $0.89$ & $4.34$ & - \\
\hline
 $a$ & 0.22 & 0.47& 0.48\\
\hline
$\sigma$ & $0.18\,\text{GeV}^2$ & $0.18\,\text{GeV}^2$ & $0.18\,\text{GeV}^2$\\
\hline
\end{tabular}
\end{center}

\caption{Comparison with \cite{Andreev:2006ct}.}
\label{AndreComp}
\end{table}

\end{center}

%%%%%%%%%%%%%%%%%%%%%%%%%%%%%%%%%%%%%%%%%%%%%%%%%%%%%%%%%%%%%%%%%%%%%%%%%%%%%%%%

\textbf{Acknowledgement}\\

We thank D. Antonov for useful discussions.

%%%%%%%%%%%%%%%%%%%%%%%%%%%%%%%%%%%%%%%%%%%%%%%%%%%%%%%%%%%%%%%%%%%%%%%%%%%%%

% ------- Bibliography -------------------------------------


\begin{thebibliography}
{}


%\cite{Randall:1999ee}
\bibitem{Randall:1999ee}
  L.~Randall and R.~Sundrum,
  %``A large mass hierarchy from a small extra dimension,''
  Phys.\ Rev.\ Lett.\  {\bf 83} (1999) 3370
  [arXiv:hep-ph/9905221].
  %%CITATION = PRLTA,83,3370;%%

%\cite{Yndurain:1999ui}
\bibitem{Yndurain:1999ui}
  F.~J.~Yndurain,
  %``The theory of quark and gluon interactions,''
%\href{/spires/find/hep/www?irn=4416570}{SPIRES entry}
{\it  Berlin, Germany: Springer (1999)}

%\cite{Aharony:1999ti}
\bibitem{Aharony:1999ti}
  O.~Aharony, S.~S.~Gubser, J.~M.~Maldacena, H.~Ooguri and Y.~Oz,
  %``Large N field theories, string theory and gravity,''
  Phys.\ Rept.\  {\bf 323} (2000) 183
  [arXiv:hep-th/9905111].
  %%CITATION = PRPLC,323,183;%%

%\cite{Kiritsis:2007zz}
\bibitem{Kiritsis:2007zz}
  E.~Kiritsis,
  %``String theory in a nutshell,''
%\href{http://www.slac.stanford.edu/spires/find/hep/www?irn=7300247}{SPIRES
%entry}
{\it  Princeton, USA: Univ. Pr. (2007)}





%\cite{Polchinski:2001tt}
\bibitem{Polchinski:2001tt}
  J.~Polchinski and M.~J.~Strassler,
  %``Hard scattering and gauge/string duality,''
  Phys.\ Rev.\ Lett.\  {\bf 88} (2002) 031601
  [arXiv:hep-th/0109174].
  %%CITATION = PRLTA,88,031601;%%

%\cite{Erlich:2005qh}
%\bibitem{Erlich:2005qh}
 % J.~Erlich, E.~Katz, D.~T.~Son and M.~A.~Stephanov,
  %``QCD and a Holographic Model of Hadrons,''
  %Phys.\ Rev.\ Lett.\  {\bf 95} (2005) 261602
  %[arXiv:hep-ph/0501128].
  %%CITATION = PRLTA,95,261602;%%



%\cite{Witten:1998qj}
\bibitem{Witten:1998qj}
  E.~Witten,
  %``Anti-de Sitter space and holography,''
  Adv.\ Theor.\ Math.\ Phys.\  {\bf 2} (1998) 253
  [arXiv:hep-th/9802150].
  %%CITATION = 00203,2,253;%%

%\cite{Gubser:1998bc}
\bibitem{Gubser:1998bc}
  S.~S.~Gubser, I.~R.~Klebanov and A.~M.~Polyakov,
  %``Gauge theory correlators from non-critical string theory,''
  Phys.\ Lett.\  B {\bf 428} (1998) 105
  [arXiv:hep-th/9802109].
  %%CITATION = PHLTA,B428,105;%%

%\cite{Maldacena:1998im}
\bibitem{Maldacena:1998im}
  J.~M.~Maldacena,
  %``Wilson loops in large N field theories,''
  Phys.\ Rev.\ Lett.\  {\bf 80} (1998) 4859
  [arXiv:hep-th/9803002].
  %%CITATION = PRLTA,80,4859;%%



\bibitem{Andreev:2006ct}
  O.~Andreev and V.~I.~Zakharov,
  %``Heavy-quark potentials and AdS/QCD,''
  Phys.\ Rev.\  D {\bf 74} (2006) 025023
  [arXiv:hep-ph/0604204].
  %%CITATION = PHRVA,D74,025023;%%

%\cite{Eichten:1978tg}
\bibitem{Eichten:1978tg}
  E.~Eichten, K.~Gottfried, T.~Kinoshita, K.~D.~Lane and T.~M.~Yan,
  %``Charmonium: The Model,''
  Phys.\ Rev.\  D {\bf 17}, 3090 (1978)
  [Erratum-ibid.\  D {\bf 21}, 313 (1980)].
  %%CITATION = PHRVA,D17,3090;%%

\bibitem{newpaper}
H. J. Pirner, B. Galow and J. Nian in preparation.

%\cite{Gursoy:2007cb}
\bibitem{Gursoy:2007cb}
  U.~Gursoy and E.~Kiritsis,
  %``Exploring improved holographic theories for QCD: Part I,''
  JHEP {\bf 0802} (2008) 032
  [arXiv:0707.1324 [hep-th]].
  %%CITATION = JHEPA,0802,032;%%

%\cite{Gursoy:2007er}
\bibitem{Gursoy:2007er}
  U.~Gursoy, E.~Kiritsis and F.~Nitti,
  %``Exploring improved holographic theories for QCD: Part II,''
  JHEP {\bf 0802} (2008) 019
  [arXiv:0707.1349 [hep-th]].
  %%CITATION = JHEPA,0802,019;%%






  %\cite{Kinar:1998vq}
\bibitem{Kinar:1998vq}
  Y.~Kinar, E.~Schreiber and J.~Sonnenschein,
  %``Q anti-Q potential from strings in curved spacetime: Classical results,''
  Nucl.\ Phys.\  B {\bf 566}, 103 (2000)
  [arXiv:hep-th/9811192].
  %%CITATION = NUPHA,B566,103;%%

%\cite{White:2007tu}
\bibitem{White:2007tu}
  C.~D.~White,
  %``The Cornell potential from general geometries in AdS / QCD,''
  Phys.\ Lett.\  B {\bf 652} (2007) 79
  [arXiv:hep-ph/0701157].
  %%CITATION = PHLTA,B652,79;%%


%\cite{Dubovsky:2001pe}
\bibitem{Dubovsky:2001pe}
  S.~L.~Dubovsky and V.~A.~Rubakov,
  %``On models of gauge field localization on a brane,''
  Int.\ J.\ Mod.\ Phys.\  A {\bf 16} (2001) 4331
  [arXiv:hep-th/0105243].
  %%CITATION = IMPAE,A16,4331;%%


\end{thebibliography}
\end{document}